\documentclass[doublespacing]{elsart}
\usepackage{graphicx,amssymb}
\journal{Physica A}

\begin{document}

\begin{frontmatter}

\title{Magnetic properties of a mixed spin-1/2 and spin-3/2 Ising model
       with an uniaxial and biaxial crystal-field potential\thanksref{VEGA}}
\author{Michal Ja\v{s}\v{c}ur} and
\ead{jascur@upjs.sk}       
\author{Jozef Stre\v{c}ka\corauthref{cor1}} 
\ead{jozkos@pobox.sk}
\address{Department of Theoretical Physics and Astrophysics, 
Faculty of Science, \\ P. J. \v{S}af\'{a}rik University, Park Angelinum 9,
041 54 Ko\v{s}ice, Slovak Republic}
\thanks[VEGA]{This work was financially supported by the VEGA 
grant No. 1/2009/05 \\ and the APVT grant No. 20-005204.}  
\corauth[cor1]{Corresponding author.}                            

\begin{abstract}
Magnetic properties of a mixed spin-1/2 and spin-3/2 Ising model
on honeycomb lattice are investigated within the framework of an
exact star-triangle mapping transformation. The particular attention
is focused on the effect of uniaxial and biaxial crystal-field potentials
that basically influence a magnetic behavior of the spin-3/2 atoms.
Our results for basic thermodynamic quantities as well as 
dynamical time-dependent autocorrelation function indicate a spin
tunneling between $| \pm \frac32 \rangle$ and $| \mp \frac12 \rangle$
states in two different magnetically ordered phases OP$_1$ and
OP$_2$, respectively.
\end{abstract}

\end{frontmatter}

\section{Introduction}

During the last decade, a quantum tunneling of magnetization
has become among the most actively studied topics in a
condensed matter physics. The immense interest aimed at better
understanding of this quantum phenomenon has been mainly 
stimulated by a recent experimental observation of the quantum
spin tunneling in a large number of single-molecule magnets
(see Ref. \cite{QT} and references therein).
By the term single-molecule magnet one denotes an assembly of 
weakly interacting clusters of magnetic metal atoms that 
usually possess an extraordinary strong magnetic anisotropy.
Hence, the single-molecule magnets often provide very good
examples of magnetic systems with a strong uniaxial magnetocrystalline
anisotropy, i.e. so-called Ising-like spin systems.
Of course, the Ising-type interaction by itself cannot
be a source of the quantum spin tunneling experimentally
observed in these systems. It turns out, however, that this quantum
phenomenon arises in the most cases due to the higher-order crystal-field
terms. According to a number of experimental and theoretical studies
it is now quite well established that the spin tunneling observed 
in Fe$_{4}$ \cite{fe4}, Fe$_8$ \cite{fe8}, Fe$_{19}$ \cite{fe19},
or Mn$_4$ \cite{mn4} clusters originates to a major extent from
a second-order biaxial crystal-field potential.

Extensive studies focused on the magnetic properties of small clusters shed 
light on the effect of single-ion anisotropy terms 
$D$ (uniaxial anisotropy) and $E$ (biaxial, also called rhombic anisotropy).
In contrast to a quite well understood role of both single-ion
anisotropies in the small magnetic clusters (zero-dimensional
systems), the situation becomes much more complex and also obscure
in one- and two-dimensional spin systems. In fact, ground-state properties
of a spin-$S$ Ising model with the rhombic crystal-field potential $E$
have been only recently examined by Oitmaa and von Brasch within
an effective mapping to the transverse Ising model \cite{oitmaa}.
On the basis of this effective mapping, a zero-temperature quantum critical
point can be exactly located in the one-dimensional model, while
for the two-dimensional models, they can be estimated with a high
numerical accuracy using the linked-cluster expansion method
\cite{oitmaa,pan}.
Nevertheless, the magnetic behavior of these models has not been
investigated at non-zero temperatures beyond the standard mean-field and
effective-field theories \cite{eddeqaqi}, random phase approximation
\cite{rpa}, or above mentioned linked cluster expansion \cite{wang}.
It should be stressed that the biaxial anisotropy
essentially influences magnetic properties of a large
number of polymeric molecular-based magnetic materials, too.
From the most obvious examples one could mention: NiF$_2$ \cite{moriya},
NiNO$_3$.6H$_2$O \cite{befr}, Ni(CH$_3$COO)$_2$.4H$_2$O \cite{pofr},
Mn(CH$_3$COO)$_2$.3H$_2$O \cite{kambik}, CoF$_2$ \cite{lines},
CoCl$_2$.6H$_2$O \cite{Uryu} and a series of compounds
$\mbox{Fe} (\mbox{dc})_2 \mbox{X}$ \cite{Wickman}, where X stands
for halides and dc for the dithiocarbamate or diselenocarbamate groups,
respectively.

In this article, we will focus on the effect of uniaxial and biaxial crystal-field 
potentials affecting the magnetic behavior of mixed spin-1/2 and spin-3/2 honeycomb lattice.
When assuming the Ising-type exchange interaction between nearest neighbors,
the model becomes exactly solvable within an extended star-triangle 
mapping transformation. Thus, the considered model provides a noble example
of statistical system, which enables to study an interplay between  
quantum effects and temperature in a spontaneously ordered magnetic system. Moreover, a magnetic structure of the mixed-spin honeycomb lattice occurs rather frequently also in the molecular magnetism, what clearly demonstrates a large family of polymeric two-dimensional compounds of following chemical formula:
$\mbox{A}^{\scriptsize \mbox{I}} \mbox{M}^{\scriptsize{\mbox{II}}}
\mbox{M}^{\scriptsize{\mbox{III}}} (\mbox{C}_2 \mbox{O}_4)_3$ \cite{car},
where A$^{\scriptsize{\mbox{I}}}$ stands for a non-magnetic univalent cation
$\mbox{N}(\mbox{C}_{n} \mbox{H}_{2n+1})_4$ or
$\mbox{P}(\mbox{C}_{n} \mbox{H}_{2n+1})_4$ ($n = 3-5$),
M$^{\scriptsize{ \mbox{II}}}$ and M$^{\scriptsize{ \mbox{III}}}$ denote
two- and three-valent metal atoms
Cu$^{\scriptsize{ \mbox{II}}} (S=1/2)$, Ni$^{\scriptsize{ \mbox{II}}} (S=1)$,
Co$^{\scriptsize{\mbox{II}}} (S=3/2)$, Fe$^{\scriptsize{\mbox{II}}} (S=2)$
or Mn$^{\scriptsize{ \mbox{II}}} (S=5/2)$ and respectively, 
Cr$^{\scriptsize{ \mbox{III}}} (S=3/2)$ or Fe$^{\scriptsize{ \mbox{III}}} (S=5/2)$. Actually, it turns out that the crystal structure of these polymeric molecular-based magnetic materials consists of the well-separated two-dimensional layers, in which regularly alternating M$^{\scriptsize{ \mbox{II}}}$ and M$^{\scriptsize{ \mbox{III}}}$ magnetic metal atoms constitute more or less regular honeycomb lattice (Fig. 1). In consequence of a large magnetocrystalline anisotropy of these  materials, one should also expect a relatively strong uniaxial (Ising-like) anisotropy, as it has already been suggested in the theoretical studies
based on the effective-field theory and Monte-Carlo simulations \cite{mc}.
Hence, the magnetic compounds from the family of oxalates
$\mbox{A}^{\scriptsize \mbox{I}} \mbox{M}^{\scriptsize{\mbox{II}}}
\mbox{M}^{\scriptsize{\mbox{III}}} (\mbox{C}_2 \mbox{O}_4)_3$
represent good candidates to be described by the proposed model.

The outline of this paper is as follows. In the next section, 
a detailed description of the model system is presented
and then, some basic aspects of the transformation method will be
shown. Section 3 deals with a physical interpretation of the
most interesting results and finally, some concluding remarks are
drawn in Section 4.

\section{Model and method}

Let us consider a magnetic structure of the mixed-spin
honeycomb lattice, as it is schematically depicted in Fig. 1.
To ensure exact tractability of the model system,
we will further suppose that the sites of sublattice $A$ are occupied
by the spin-1/2 atoms (depicted as full circles), in contrast to the sites
of sublattice $B$ that are occupied by the spin-3/2 atoms (open circles).
Assuming the Ising-type exchange interaction $J$ between nearest-neighboring 
spin pairs only, the total Hamiltonian of the system reads:
\begin{equation}
\hat {\mathcal H} = J \sum_{\langle k, j \rangle}^{3N} \hat S_k^z \hat \mu_j^z
              + D \sum_{k \in B}^{N} (\hat S_k^z)^2
              + E \sum_{k \in B}^{N} [(\hat S_k^x)^2 - (\hat S_k^y)^2],
\label{eq1}
\end{equation}
where $N$ is a total number of sites at each sublattice,
$\hat \mu_j^z$ and $\hat S_k^{\alpha} (\alpha = x, y, z)$
denote standard spatial components of the spin-1/2 and spin-3/2
operators, respectively. The first summation in Eq. (\ref{eq1})
is carried out over nearest-neighboring spin pairs, while
the other two summations run over the sites of sublattice $B$.
Apparently, the last two terms $D$ and $E$ are the crystal-field
potentials that measure a strength of uniaxial and biaxial
anisotropies acting on the spin-3/2 atoms. It is also worthy to note
that there is one-to-one correspondence between the Hamiltonian (\ref{eq1})
and the effective spin Hamiltonian with three different
single-ion anisotropy terms $D^x$, $D^y$ and $D^z$:
\begin{equation}
\hat {\mathcal H} = J \sum_{\langle k, j \rangle}^{3N} \hat S_k^z \hat \mu_j^z
              + D^z \sum_{k \in B}^{N} (\hat S_k^z)^2
              + D^x \sum_{k \in B}^{N} (\hat S_k^x)^2
              + D^y \sum_{k \in B}^{N} (\hat S_k^y)^2.
\label{eq2}
\end{equation}
As a matter of fact, one can easily prove the equivalence between
(\ref{eq1}) and (\ref{eq2}) using following mapping 
relations between the relevant interaction parameters:
\begin{equation}
D = D^z - \frac12 (D^x + D^y), \qquad \mbox{and} \qquad E = \frac12 (D^x - D^y).
\label{eq3}
\end{equation}
It should be also mentioned here that by neglecting the biaxial
anisotropy, i.e. setting $E = 0$ in Eq. (\ref{eq1})
or equivalently $D^x = D^y$ in Eq. (\ref{eq2}),
our model reduces to the exactly soluble model settled by Gon\c{c}alves \cite{bc} 
several years ago. Accordingly, the main attention will be focused here on the effect of
biaxial anisotropy, which influences thermodynamical and dynamical properties
in a crucial manner. Really, the $E$-term related to the
biaxial crystal-field anisotropy should cause non-trivial quantum effects,
since it introduces the $x$ and $y$ components of spin operators into
the Hamiltonian (\ref{eq1}). As a result,
it is responsible for the onset of local quantum fluctuations that
are obviously missing in the Ising model with the uniaxial crystal-field
potential $D$ only.

It is therefore of interest to discuss an origin of the biaxial
crystal-field potential $E$.
The origin of this anisotropy term consists in the low-symmetry crystal field
of ligands from the local neighborhood of the spin-3/2 atoms. It is
noticeable that a threefold symmetry
axis oriented perpendicular to the honeycomb layer prevents
an appearance of biaxial crystal-field potential in a regular honeycomb
lattice with a perfect arrangement of the oxalato groups, as
well as magnetic metal atoms. However, a small lattice distortion which occurs 
rather frequently in the low-dimensional polymeric compounds
due to the Jahn-Teller effect can potentially lower the local symmetry. In
consequence of that, the distortion of lattice parameters can be regarded
as a possible source of the biaxial crystal-field anisotropy. The most obvious
example, where the lattice distortion removes the threefold
symmetry axis represents the single-molecule magnet Fe$_4$,
in which three outer Fe atoms occupy two non-equivalent
positions around one central Fe atom \cite{fe4}.

Let us turn our attention to the main points of the
transformation method, which enables an exact treatment of the model
system. Firstly, it is very convenient to write the total Hamiltonian
(\ref{eq1}) as a sum of site Hamiltonians:  
\begin{equation}
\hat {\mathcal H} = \sum_{k \in B}^{N} \hat {\mathcal H}^{(k)},
\label{eq4}
\end{equation}
where the each site Hamiltonian $\hat {\mathcal H}^{(k)}$ involves all interaction
terms associated with one spin-3/2 atom residing on the $k$th
site of sublattice $B$:
\begin{equation}
\hat {\mathcal H}^{(k)} = \hat S_{k}^{z} E_k
                  + (\hat S_{k}^{z})^2 D
                  + [(\hat S_{k}^{x})^2 - (\hat S_{k}^{y})^2] E,
\label{eq5}
\end{equation}
with $E_k  =  J (\hat \mu_{k1}^{z} + \hat \mu_{k2}^{z} + \hat \mu_{k3}^{z})$.
While the Hamiltonians (\ref{eq5}) at different sites commute with respect to 
each other ($[\hat {\mathcal H}^{(i)}, \hat {\mathcal H}^{(j)}] = 0$, for each $i \neq j$), the partition function of the system can be partially factorized
and consequently, rewritten in the form:
\begin{equation}
{\mathcal Z} = \displaystyle{\mbox{Tr}_{\{\mu \}}}
\prod_{k = 1}^{N} \mbox{Tr}_{S_k} \exp[- \beta \hat {\mathcal H}^{(k)}].
\label{eq6}
\end{equation}
In above, $\beta = 1/(k_B T)$, $k_B$ is Boltzmann's constant,
$T$ the absolute temperature, $\mbox{Tr}_{ \{ \mu \} }$ means
a trace over  spin degrees of freedom of sublattice $A$ and
$\mbox{Tr}_{S_k}$ stands for a trace over spin states
of the $k$th spin from sublattice $B$. A crucial step in our
procedure represents the calculation of the expression
$\mbox{Tr}_{S_k} \exp[- \beta \hat {\mathcal H}^{(k)}]$. With regard
to this, let us write the site Hamiltonian (\ref{eq5}) in an usual
matrix representation:
\begin{eqnarray}
\hat {\mathcal H}^{(k)}  =
\left(
\begin{array}{cccc}
\displaystyle \frac{9D}{4} + \displaystyle \frac{3E_k}{2} &  0  & \sqrt{3}E &   0 \\
  0     &  \displaystyle \frac{D}{4} + \displaystyle \frac{E_k}{2}  &   0  &  \sqrt{3}E  \\
\sqrt{3}E   &  0  & \displaystyle \frac{D}{4} - \displaystyle \frac{E_k}{2} &   0 \\
  0     &  \sqrt{3}E  &   0  &  \displaystyle \frac{9D}{4} - \displaystyle \frac{3E_k}{2}   \\
\end{array}
                  \right),
\label{eq7}
\end{eqnarray}
in a standard basis of functions $| \pm 3/2 \rangle, | \pm 1/2 \rangle$
corresponding, respectively, to the four possible spin states
$S_k^z = \pm 3/2, \pm 1/2$ of the $k$th atom from sublattice $B$.
Although it is easy to find eigenvalues of the site Hamiltonian
(\ref{eq7}), with respect to further calculation
it is more favorable to obtain directly matrix elements of the 
expression $\exp[- \beta \hat {\mathcal H}^{(k)}]$. When adopting 
the Cauchy integral formula, one readily attains the matrix elements for
an arbitrary exponential function of the site Hamiltonian (\ref{eq7}):
\begin{eqnarray}
A_{ij} &=& 
\Bigl( \exp [\alpha \hat {\mathcal H}^{(k)}] \Bigr)_{ij},
\nonumber \\
A_{11} &=& 
\exp \Bigl[\alpha (\frac54 D + \frac12 E_k) \Bigr] \Bigl \{
\cosh(a \xi_k^{+}) + \frac{D + E_k}{\xi_k^{+}} \sinh(a \xi_k^{+}) \Bigr \},
\nonumber \\
A_{22} &=& 
\exp \Bigl[\alpha (\frac54 D - \frac12 E_k) \Bigr] \Bigl \{
\cosh(a \xi_k^{-}) - \frac{D - E_k}{\xi_k^{-}} \sinh(a \xi_k^{-}) \Bigr \},
\nonumber \\
A_{33} &=& 
\exp \Bigl[\alpha (\frac54 D + \frac12 E_k) \Bigr] \Bigl \{
\cosh(a \xi_k^{+}) - \frac{D + E_k}{\xi_k^{+}} \sinh(a \xi_k^{+}) \Bigr \},
\nonumber \\
A_{44} &=& 
\exp \Bigl[\alpha (\frac54 D - \frac12 E_k) \Bigr] \Bigl \{
\cosh(a \xi_k^{-}) + \frac{D - E_k}{\xi_k^{-}} \sinh(a \xi_k^{-}) \Bigr \},
\nonumber \\
A_{13} &=& A_{31} =
\exp \Bigl[\alpha (\frac54 D + \frac12 E_k) \Bigr]
\frac{\sqrt{3} E}{\xi_k^{+}} \sinh(a \xi_k^{+}),
\nonumber \\
A_{24} &=&  A_{42} =
\exp \Bigl[\alpha (\frac54 D - \frac12 E_k) \Bigr]
\frac{\sqrt{3} E}{\xi_k^{-}} \sinh(a \xi_k^{-}),
\label{eq8}
\end{eqnarray}
where $\xi_k^{\pm} = \sqrt{(D \pm E_k)^2 + 3 E^2}$ and $\alpha$ marks
an arbitrary function. After substituting $\alpha = - \beta$
in the set of Eqs. (\ref{eq8}), the relevant trace 
$\mbox{Tr}_{S_k} \exp[- \beta \hat {\mathcal H}^{(k)}]$ can easily
be calculated. Moreover, its explicit form immediately implies
a possibility of performing the standard star-triangle mapping
transformation:
\begin{eqnarray}
\mbox{Tr}_{S_k} \exp [&-& \beta \hat {\mathcal H}^{(k)}] = 
2 \exp[-5 \beta D/4 - \beta E_{k}/2]
\cosh \Bigl( \beta \sqrt{(D + E_{k})^2 + 3 E^2} \Bigr) 
\nonumber \\
&+& 2 \exp[-5 \beta D/4 + \beta E_{k}/2]
\cosh \Bigl( \beta \sqrt{(D - E_{k})^2 + 3 E^2} \Bigr) =
\nonumber \\
&=& A~\exp \Bigl[ \beta R (\mu_{k1}^{z} \mu_{k2}^{z} + \mu_{k2}^{z}
\mu_{k3}^{z} + \mu_{k3}^{z} \mu_{k1}^{z} \bigr) \Bigr],
\label{eq9}
\end{eqnarray}
which replaces the partition function of a {\it star} (i.e. the
four-spin cluster consisting of one central spin-3/2 atom and its three
nearest-neighboring spin-1/2 atoms) by the partition function of
a {\it triangle} (i.e. the three-spin cluster comprising of three
outer spin-1/2 atoms in the corners of equilateral triangle), as shown 
in Fig. 1. A physical meaning of the mapping (\ref{eq9}) is to
remove all interaction parameters associated with the central spin-3/2
atom and to replace them by the effective interaction $R$ between
the outer spin-1/2 atoms. It is noteworthy that both mapping
parameters $A$ and $R$ are "self-consistently" given by the transformation
equation (\ref{eq9}), which must be valid for any combination of spin
states of three spin-1/2 atoms. In consequence of that, the
transformation parameters $A$ and $R$ can be expressed as:
\begin{eqnarray}
A~= \Bigl(\Phi_{1} \Phi_{2}^3 \Bigr)^{1/4}, \qquad \qquad
\beta R = \mbox{ln} \Bigl( \frac{\Phi_1}{\Phi_2} \Bigr),
\label{eq10}
\end{eqnarray}
where the functions $\Phi_1$ and $\Phi_2$ are defined as follows:
\begin{eqnarray}
\Phi_{1} &=& 2 \exp(- 5 \beta D/4 - 3 \beta J/4) \cosh \Bigl( \beta
\sqrt{\bigl(3J/2 + D)^{2} + 3E^2} \Bigr)  \nonumber \\
&+& 2 \exp(- 5 \beta D/4 + 3 \beta J/4)
\cosh \Bigl( \beta \sqrt{\bigl(3J/2 - D)^{2} + 3E^2} \Bigr),    \nonumber \\
\Phi_{2} &=& 2 \exp(- 5 \beta D/4 - \beta J/4)
\cosh \Bigl( \beta
\sqrt{\bigl(J/2 + D)^{2} + 3E^2} \Bigr) \nonumber \\
&+& 2 \exp(- 5 \beta D/4 + \beta J/4)
\cosh \Bigl( \beta \sqrt{\bigl(J/2 - D)^{2} + 3E^2} \Bigr).
\label{eq11}
\end{eqnarray}

When the mapping (\ref{eq9}) is performed at each site of
the sublattice $B$, the original mixed-spin honeycomb lattice is
mapped onto the spin-1/2 Ising triangular lattice with the effective
interaction $R$ given by the "self-consistency" condition
(\ref{eq10})-(\ref{eq11}). Indeed, the substitution
of equation (\ref{eq9}) into the partition
function (\ref{eq6}) establishes the relationship:
\begin{equation}
{\mathcal Z} (\beta, J, D, E) = A^{N} {\mathcal Z}_{t} (\beta, R),
\label{eq12}
\end{equation}
between the partition function ${\mathcal Z}$ of the mixed-spin honeycomb
lattice and the partition function ${\mathcal Z}_t$ of the corresponding
spin-1/2 triangular lattice. Above equation
constitutes the basic result of our calculation, since it enables
relatively simple derivation of all required quantities, such as the
magnetization, quadrupolar moment, correlation function,
internal energy, specific heat, etc. In addition, by combining 
(\ref{eq12}) with (\ref{eq9}) one readily proves a validity of
following exact spin identities:
\begin{eqnarray}
\langle f_1 (\mu_i^z, \mu_j^z, ..., \mu_k^z) \rangle
&=& \langle f_1 (\mu_i^z, \mu_j^z, ..., \mu_k^z) \rangle_t,
\label{eq13a} \\
\langle f_2 (S_k^x, S_k^y, S^z_k, \! \! && \! \! \mu_{k1}^z, \mu_{k2}^z, \mu_{k3}^z) \rangle = \nonumber \\
&=& \Biggl \langle
\frac{\mbox{Tr}_{S_k} f_2 (S_k^x, S_k^y, S^z_k, \mu_{k1}^z, \mu_{k2}^z, \mu_{k3}^z)
\exp[- \beta \hat {\mathcal H}^{(k)}]}{\mbox{Tr}_{S_k} \exp[- \beta \hat 
{\mathcal H}^{(k)}]}
\Biggr \rangle,
\label{eq13}
\end{eqnarray}
where $\langle ... \rangle$ represents the standard canonical
average over the ensemble defined by the Hamiltonian (\ref{eq1}) and
$\langle ... \rangle_t$ the one performed on the spin-1/2
Ising triangular lattice with the effective exchange interaction $R$.
Here, $f_1$ is an arbitrary function of the spin variables
belonging to the sublattice $A$, while $f_2$ denotes an arbitrary
function depending on the $k$th spin of sublattice $B$ and
its three nearest neighbors from the sublattice $A$. By applying the
spin identity (\ref{eq13a}), one straightforwardly
attains the following results:
\begin{eqnarray}
m_A &\equiv& \langle \hat \mu_{k1}^z \rangle
    = \langle \hat \mu_{k1}^z \rangle_t \equiv m_t, \\
c_A &\equiv& \langle \hat \mu_{k1}^z \hat \mu_{k2}^z \rangle
      = \langle \hat \mu_{k1}^z \hat \mu_{k2}^z \rangle_t \equiv c_t,  \\
t_A &\equiv& \langle \hat \mu_{k1}^z \hat \mu_{k2}^z \hat \mu_{k3}^z \rangle
  = \langle \hat \mu_{k1}^z \hat \mu_{k2}^z \hat \mu_{k3}^z  \rangle_t
\equiv t_t,
\label{eq14}
\end{eqnarray}
whereas the second spin identity (\ref{eq13}) enables after
some algebra derivation of quantities depending on the spin variable
from the sublattice $B$:
\begin{eqnarray}
m_B &\equiv& \langle \hat S_k^z \rangle =
- 3 m_A (K_1 + K_2)/2 - 2 t_A (K_1 - 3 K_2),  \\
\eta &\equiv& \langle (\hat S_k^z)^2 \rangle =
(K_3 + 3 K_4)/4 + 3 c_A (K_3 - K_4).
\label{eq145}
\end{eqnarray}
In above, $m_A$ ($m_B$) labels the single-site magnetization of the
sublattice $A$ ($B$), $\eta$ denotes the quadrupolar moment
and finally, $c_A$ and $t_A$ stand, respectively, for the static pair 
and triplet correlation functions between the relevant spins of sublattice 
$A$. Obviously, the exact solution of both sublattice
magnetization and quadrupolar moment require only a knowledge
of the single-site magnetization $m_t$, nearest-neighbour pair
correlation function $c_t$ and triplet correlation function
$t_t$ on the corresponding spin-1/2 Ising triangular lattice unambiguously
given by $R$ (\ref{eq10})-(\ref{eq11}). Fortunately, the appropriate exact
solution of these quantities is well known, hence, one can
directly utilize final results derived in Refs. \cite{pot}. Finally,
the coefficients emerging in the previous set of Eqs. (17)-(20) are listed
below:
\begin{eqnarray}
K_1 &=& F_1 (3J/2 + D, 3J/2 - D, 3J/2),  \hspace{2mm}
K_2 = F_1 (J/2 + D, J/2 - D, J/2), \nonumber \\
K_3 &=& F_2 (3J/2 + D, 3J/2 - D, 3J/2), \hspace{2mm}
K_4 = F_2 (J/2 + D, J/2 - D, J/2), \nonumber \\
\label{eq15}
\end{eqnarray}
where the functions $F_1 (x, y, z)$ and $F_2 (x, y, z)$ are
defined as follows:
\begin{eqnarray}
F_1 (&x&, y, z) = \frac{x}{\sqrt{x^2 + 3E^2}}
\frac{\sinh(\beta \sqrt{x^2 + 3E^2})}
     {\cosh(\beta \sqrt{x^2 + 3E^2})
    + \exp(\beta z) \cosh(\beta \sqrt{y^2 + 3E^2})} \nonumber \\
&+& \frac{y}{\sqrt{y^2 + 3E^2}}
\frac{\sinh(\beta \sqrt{y^2 + 3E^2})}
     {\exp(- \beta z) \cosh(\beta \sqrt{x^2 + 3E^2})
    + \cosh(\beta \sqrt{y^2 + 3E^2})}  \nonumber \\
&-& \frac12 \frac{\cosh(\beta \sqrt{x^2 + 3E^2}) -
             \exp(\beta z) \cosh(\beta \sqrt{y^2 + 3E^2})}
     {\cosh(\beta \sqrt{x^2 + 3E^2}) +
            \exp(\beta z) \cosh(\beta \sqrt{y^2 + 3E^2})};  \nonumber \\
F_2 (&x&, y, z) = \frac{y}{\sqrt{y^2 + 3E^2}}
\frac{\sinh(\beta \sqrt{y^2 + 3E^2})}
     {\exp(- \beta z) \cosh(\beta \sqrt{x^2 + 3E^2})
    + \cosh(\beta \sqrt{y^2 + 3E^2})} \nonumber \\
 &+& \frac54 - \frac{x}{\sqrt{x^2 + 3E^2}}
\frac{\sinh(\beta \sqrt{x^2 + 3E^2})}
     {\cosh(\beta \sqrt{x^2 + 3E^2})
    + \exp(\beta z) \cosh(\beta \sqrt{y^2 + 3E^2})}. \nonumber \\
\label{eq155}
\end{eqnarray}

Now, we will derive an exact result for one dynamical 
quantity, namely, the time-dependent autocorrelation function.
It should be noted here that exactly tractable models offer only seldom a
possibility to investigate their spin dynamics. On the other hand, the dynamical
quantities such as autocorrelation and correlation functions are important
also from the experimental point of view, because their magnitude
directly determines a scattering cross section measured in inelastic
neutron scattering experiments \cite{nse}, or a spin-lattice relaxation
rate provided by a nuclear magnetic resonance technique \cite{nmr}.

As a starting point for calculation of
the time-dependent autocorrelation function $C_{auto}^{zz}$ can, for
convenience, serve the exact spin identity (\ref{eq13}):
\begin{eqnarray}
C_{auto}^{zz} (t) &\equiv& \frac12 \langle
\hat S^z_k (0) \hat S^z_k (t) + \hat S^z_k (t) \hat S^z_k (0)
\rangle = \nonumber \\
&=& \frac12 \biggl \langle
\frac{\mbox{Tr}_{S_k} \{ [\hat S^z_k (0) \hat S^z_k (t) +
\hat S^z_k (t) \hat S^z_k (0)] \exp[- \beta \hat {\mathcal H}^{(k)}] \}}
     {\mbox{Tr}_{S_k} \exp[- \beta \hat {\mathcal H}^{(k)}]}
\biggr \rangle,
\label{eq16}
\end{eqnarray}
where the symmetrized form in the definition of $C_{auto}^{zz}$
is used to construct a Hermitian operator,
$\hat S^z_k (t) = \exp(\frac{i t \hat {\mathcal H}_k}{\hbar}) \hat S^z_k \exp(- \frac{i t \hat {\mathcal H}_k}{\hbar})$ represents the Heisenberg picture for the time-dependent operator $\hat S^z_k (t)$,
$\hbar$ stands for the reduced Planck's constant and $i = \sqrt{-1}$.
Next, the matrix elements of expressions 
$\exp(\pm \frac{i t \hat {\mathcal H}_k}{\hbar})$
can be in turn evaluated by putting $\alpha = \pm \frac{i t}{\hbar}$ into
the set of Eqs. (\ref{eq8}). Then, after a straightforward but a little bit tedious
calculation, one arrives to a final result for the dynamical
autocorrelation function:
\begin{eqnarray}
&C&_{auto}^{zz} = 
\frac14 [K_5(D + 3J/2, D - 3J/2, 3J/2) + 3K_6(D + J/2, D - J/2, J/2)]
\nonumber \\ &+& 
3 c_t [K_5(D + 3J/2, D - 3J/2, 3J/2) - K_6(D + J/2, D - J/2, J/2)],
\label{eq17}
\end{eqnarray}
where the time-dependent coefficients $K_5$ and $K_6$ are, 
for brevity, explicitly given in the Appendix.

\section{Results and discussion}

Before proceeding to a discussion of the most interesting
results, it is noticeable that the results derived in the
previous section are rather general, since they are valid for
a ferromagnetic ($J < 0$) as well as ferrimagnetic ($J > 0$)
version of the model under investigation. In what follows,
we shall restrict our analysis to the ferrimagnetic
model only, since the polymeric compounds from the family of
oxalates \cite{car} fall mostly into the class of ferrimagnets.
Nevertheless, it appears worthwhile to remark that a magnetic
behavior of the ferrimagnetic system completely resembles
the one of ferromagnetic system.
Finally, one should also emphasize that the mapping (\ref{eq9})
remains invariant under the transformation $E \leftrightarrow -E$.
For this reason, one may consider without loss of generality the parameter
$E \geq 0$ and consequently, $x$-, $y$- and $z$-axis  then 
represent the hard-, medium- and easy-axis for a given system.

\subsection{Ground-state properties}

At first, we will take a closer look at the ground-state
behaviour. Taking into account the zero-temperature
limit ($T \to 0^{+}$), one finds the following condition
for a first-order phase transition line separating
two different magnetically ordered phases denoted as OP$_1$ and
OP$_2$:
\begin{equation}
\frac{D}{J} = \sqrt{\Bigl( \frac34 \Bigr)^2 + \Bigl( \frac{E}{J} \Bigr)^2}.
\label{eq175}
\end{equation}
Moreover, one easily attains from Eqs. (15)-(\ref{eq155})
analytical results for the single-site sublattice magnetization
($m_A$, $m_B$), the total magnetization normalized per one
magnetic atom $m = (m_A + m_B)/2$ and the quadrupolar moment
$\eta$, as well:
\begin{eqnarray}
\mbox{OP$_1$:} \hspace*{0.25cm} m_A &=& - \frac12,
\hspace*{0.9cm}
m_B = \frac12  +  \frac{\frac32 - \frac{D}{J}}
           {\sqrt{\Bigl(\frac32 - \frac{D}{J} \Bigr)^2
                  + 3 \Bigl( \frac{E}{J} \Bigr)^2}},  \\
m &=&  \frac12 \frac{\frac32 - \frac{D}{J}}
           {\sqrt{\Bigl(\frac32 - \frac{D}{J} \Bigr)^2
                  + 3 \Bigl( \frac{E}{J} \Bigr)^2}}, \hspace*{2mm}
\eta = \frac54 + \frac{\frac32 - \frac{D}{J}}
           {\sqrt{\Bigl(\frac32 - \frac{D}{J} \Bigr)^2
                  + 3 \Bigl( \frac{E}{J} \Bigr)^2}}; \nonumber
\label{eq18}
\end{eqnarray}
\begin{eqnarray}
\mbox{OP$_2$:} \hspace*{0.25cm} m_A &=&  -\frac12,
\hspace*{0.9cm}
m_B = - \frac12  +  \frac{\frac32 + \frac{D}{J}}
           {\sqrt{\Bigl(\frac32 + \frac{D}{J} \Bigr)^2
                  + 3 \Bigl( \frac{E}{J} \Bigr)^2}},  \\
m &=&  - \frac12  - \frac12 \frac{\frac32 - \frac{D}{J}}
           {\sqrt{\Bigl(\frac32 - \frac{D}{J} \Bigr)^2
                  + 3 \Bigl( \frac{E}{J} \Bigr)^2}}, \hspace*{2mm}
\eta = \frac54 - \frac{\frac32 + \frac{D}{J}}
           {\sqrt{\Bigl(\frac32 + \frac{D}{J} \Bigr)^2
                  + 3 \Bigl( \frac{E}{J} \Bigr)^2}}. \nonumber
\label{eq19}
\end{eqnarray}
For better illustration, Fig. 2 depicts the ground-state phase
diagram in the $E$-$D$ plane (Fig. 2a) and the zero-temperature
variations of the magnetization and quadrupolar moment with
the biaxial anisotropy when $D/J = 1.0$ (Fig. 2b).
It should be mentioned that by neglecting the biaxial anisotropy,
i.e. by putting $E/J = 0.0$ into the phase boundary condition (\ref{eq175}),
one recovers the boundary uniaxial anisotropy $D/J = 0.75$ in accordance 
with the results reported by Gon\c{c}alves \cite{bc} several years ago.
Moreover, the OP$_1$ (OP$_2$) phase corresponds in this limit to the simple
ferrimagnetic (antiferromagnetic) phase with both sublattice magnetization
oriented antiparallel with respect to each other: $m_A = -0.5$,
$m_B = 1.5$ in the OP$_1$ and respectively, $m_A = -0.5$, $m_B = 0.5$
in the OP$_2$. Apparently, the spin-3/2 atoms occupy exclusively
the $|+ 3/2 \rangle$ ($|+ 1/2 \rangle$) state in the OP$_1$ (OP$_2$)
phase when $E = 0$ is satisfied.

The situation becomes much more complex by turning on the biaxial
anisotropy. Even though the sublattice magnetization $m_A$ remains
in the ground-state at its saturation value in both OP$_1$ and OP$_2$
phases, the sublattice magnetization $m_B$ is gradually suppressed by
the effect of biaxial anisotropy (see Fig. 2b). It is quite obvious from
this figure that the biaxial anisotropy gradually destroys the perfect ferrimagnetic
(antiferromagnetic) spin arrangement, which occurs in the OP$_1$ (OP$_2$)
phase in the limit of vanishing $E$. Let us find a primary occasion
for this unexpected behavior accompanied by a spin reduction at
sublattice $B$. According to Eq. (\ref{eq18}), one finds $\eta - m_B = 3/4$ to
be valid in the whole parameter space corresponding to the OP$_1$. From an elementary
consideration it can be easily understood that the spin-3/2 atoms must
occupy in the OP$_1$ phase either the $|+3/2 \rangle$ or $|-1/2 \rangle$ state
in order to satisfy simultaneously both values of $m_B$ and $\eta$ in the ground state.
Aforementioned argument is also supported by the fact that
the $E$-term does not couple in the Hamiltonian (\ref{eq7})
the $|+3/2 \rangle$ and $|-1/2 \rangle$ states with the
$|+1/2 \rangle$ and $|-3/2 \rangle$ ones. Hence,
the $|+3/2 \rangle$ and $|-1/2 \rangle$ states are in the OP$_1$ phase
the only {\it allowable} states, while the $|+1/2 \rangle$ and
$|-3/2 \rangle$ states are, on the contrary, the {\it forbidden} ones. The observed
spin reduction at sublattice $B$ can be thus attributed to the local quantum
fluctuations induced by the biaxial anisotropy, which in turn
lead to a spin tunneling between the $|+3/2 \rangle$ and $|-1/2 \rangle$
states in the OP$_1$. Although the occupation of the minority $|-1/2 \rangle$ 
state rises steadily with increasing the biaxial anisotropy strength,
it is noticeable that the $|+3/2 \rangle$ state still
remains the state with the most probable occupation.
For completeness, it should be pointed out that in accord with our
expectation the negative uniaxial anisotropy ($D < 0$) reduces
the occupation of the minority $|-1/2 \rangle$
state, while the positive one ($D > 0$) favors it.

Quite similar situation emerges also in the OP$_2$ phase.
However, it is worthwhile to remark that the OP$_2$ phase appears 
in the region of strong uniaxial anisotropies $D/J > 0.75$ only. 
Due to a strong positive uniaxial anisotropy, the spin-3/2 atoms
undergo a well-known spin transition from the $|+3/2 \rangle$
to $|+1/2 \rangle$ state, which macroscopically manifests itself 
in the phase transition from the OP$_1$ to OP$_2$ phase. As stated before, 
the biaxial anisotropy couples together the $|+1/2 \rangle$ and $|-3/2 \rangle$ 
states and therefore, the tunneling between these spin states should be
expected to occur in the OP$_2$. The analytical solution for
$m_B$ and $\eta$ (\ref{eq19}), as well as a validity of the
relation $\eta + m_B = 3/4$ in a whole parameter space corresponding 
to the OP$_2$ phase, indeed confirm this suggestion.
However, the negative (positive) uniaxial anisotropy prefers (reduces) 
the occupation of minority $| - 3/2 \rangle$ state in the OP$_2$ phase 
in contrast to the abovementioned trends observed in the OP$_1$ phase.

Now, let us step forward to the discussion of the time dependent
autocorrelation function (\ref{eq17}). Among other matters, this quantity can
serve in evidence whether the spin-3/2 atoms fluctuate in the OP$_1$
(OP$_2$) phase between their {\it allowable} $| + 3/2 \rangle$ and
$| - 1/2 \rangle$ ($| + 1/2 \rangle$ and $| - 3/2 \rangle$) spin states.
Unfortunately, it is quite tedious to derive
from Eq. (\ref{eq17}) a simple analytical expression for $C_{auto}^{zz}$
in the zero-temperature limit, hence, we report for $C_{auto}^{zz}$
numerical results obtained at very low temperature ($k_B T/J = 0.001$)
close to the ground state. Fig. 3 displays the time variation of $C_{auto}^{zz}$
for several values of uniaxial and biaxial crystal-field potentials.
Since $C_{auto}^{zz}$ evidently varies in time, it clearly demonstrates
the zero-temperature spin dynamics between the {\it allowable} states.
Moreover, a detailed analysis reveals that $C_{auto}^{zz}$ is in the zero-temperature 
limit a harmonic function of time and whence,
the time dependence can be characterized by an angular frequency
$\omega_{\pm} = \frac{2 J}{\hbar} \sqrt{(\frac{D}{J} \pm 1) + 3 (\frac{E}{J})^2}$
depending on whether the system resides the OP$_1$, or OP$_2$ phase.
This result is taken to mean that the spin system necessarily recovers 
after some characteristic recurrence time ($\tau_{\pm} = 2 \pi / \omega_{\pm}$)
its initial state. More specifically, Fig. 3a illustrates the time
variation of $C_{auto}^{zz}$ in the OP$_1$ phase, because with respect
to Eq. (\ref{eq175}) one never approaches the OP$_2$ phase for $D/J = 0.0$.
Fig. 3a clearly clarifies the role of biaxial anisotropy:
the stronger the biaxial anisotropy $E/J$, the greater the angular 
frequency of spin tunneling and in the consequence of that,
the shorter the appropriate recurrence time. Furthermore, the increasing
strength of the biaxial anisotropy enhances also the amplitude of oscillation 
in the time-dependence of $C_{auto}^{zz}$.
This observation would suggest that the increase of biaxial anisotropy
enhances a number of the spin-3/2 atoms tunneling during
the recurrence time between the $|+3/2 \rangle$ and $|-1/2 \rangle$ states
in the OP$_1$ phase. However, since the equilibrium magnetization does not 
change in time, the number of atoms that tunnel from $| +3/2 \rangle$ to $| -1/2 \rangle$ state must definitely be the same as those that tunnel from the $| -1/2 \rangle$ to $| +3/2 \rangle$ state.

To illustrate the effect of uniaxial anisotropy, the time variation of
$C_{auto}^{zz}$ is shown in Fig. 3b for $E/J = 0.5$ and several values of $D/J$.
Apparently, the $C_{auto}^{zz}$ oscillates for strong negative (positive)
uniaxial constants $D/J$ in the vicinity of boundary values 
$C_{auto}^{zz} = 2.25 (0.25)$.
These values clearly demonstrate that the prevailing
number of spin-3/2 atoms occupy in the OP$_1$ (OP$_2$) phase the
$| +3/2 \rangle$ ($| +1/2 \rangle$) state, since $C_{auto}^{zz} = \eta$
when $t = 0$. Moreover,
the stronger the uniaxial anisotropy (independently of its sign),
the smaller the relevant amplitudes of oscillation, i.e. the smaller
the number of tunneling atoms during the recurrence time. On the
other hand, the increasing strength of uniaxial anisotropy enhances
the angular frequency of oscillation, what means, that the tunneling
atoms return from the minority $| -1/2 \rangle$ ($| -3/2 \rangle$) state
to the most probable occupied $| +3/2 \rangle$ ($| +1/2 \rangle$) state 
of the OP$_1$ (OP$_2$) phase after a shorter recurrence time.

\subsection{Finite-temperature behaviour}

In this part, we would like to comment on the finite-temperature behaviour 
of the system under investigation.
Let us begin by considering the effect of uniaxial and biaxial single-ion anisotropies 
on the critical behaviour. For this purpose, two typical finite-temperature
phase diagrams are illustrated in Fig. 4a and 4b. In both figures,
the OP$_1$ (OP$_2$) phase can be located below the phase boundaries
depicted as solid (dashed) lines, while above these boundary lines
the usual paramagnetic phase becomes stable. Further, open circles
represent special critical points at which both OP$_1$ and OP$_2$ 
phases coexist. Actually, we have not found any phase
transition between the OP$_1$ and OP$_2$ phases at non-zero temperatures,
what indicates that the OP$_1$ phase coexists with the OP$_2$ one at
non-zero temperatures merely for the same $D/J-E/J$ values as in the ground
state (\ref{eq175}). Finally, a closer mathematical analysis reveals that
both temperature-driven phase transitions which are related to the OP$_1$
and OP$_2$ phase, respectively, are of a second order and belong to
a standard Ising universality class.

Fig. 4a shows the critical temperature ($T_c$) as a function of the uniaxial
anisotropy for several values of the biaxial anisotropy.
The critical temperature versus uniaxial anisotropy dependence 
is quite obvious for $E/J = 0.0$,
when increasing $D/J$, $T_c$ monotonically decreases.
Hence, the critical temperature approaches in the limit
$D/J \to - \infty (+ \infty)$ its minimum (maximum) value
$k_B T_c /J = 0.3796...$ ($1.1389...$) in agreement with the
exact $T_c$ (triple $T_c$) of the spin-1/2 Ising honeycomb lattice.
A gradual decline of the transition temperature can obviously be  
explained as a consequence of the fact, that the positive uniaxial anisotropy
energetically favors the low-spin $| \pm 1/2 \rangle$ states before the 
high-spin $| \pm 3/2 \rangle$ ones. 
The most interesting finding to emerge here is that the biaxial
anisotropy may significantly modify the critical behavior of the studied
system. As a matter of fact, $T_c$ firstly reaches its local minimum at certain 
positive $D/J$ and then rises steadily to its limiting value $k_B T_c /J = 0.3796...$. 
The extraordinary increase of $T_c$ in the region $D/J > 1.0$
can be explained through a suppression of the occupation of
minority $| -3/2 \rangle$ state, which appears in the OP$_2$ phase
due to the uniaxial anisotropy effect.
In accordance with previous assumption, the greater the biaxial
anisotropy (i.e. the greater a number of atoms that occupy the
minority $| -3/2 \rangle$ state), the more impressive increase of $T_c$ can
be observed. In addition, it is easy to understand from here that the biaxial
anisotropy substantially lowers the critical temperature of OP$_1$ phase
in the $D/J \leq 0.0$ region, in that it is responsible for the quantum
spin tunneling between the $| +3/2 \rangle$ and $| -1/2 \rangle$ states.

To illustrate the influence of biaxial anisotropy on the criticality, 
the dependence of transition temperature on the biaxial anisotropy 
is shown in Fig. 4b for several values of the uniaxial anisotropy $D/J$. 
As one would expect,
$T_c$ gradually decreases with increasing the biaxial anisotropy
for any $D/J < 0.75$. It is quite obvious that the appropriate 
suppression of $T_c$ can be attributed to the
quantum fluctuations, which become the stronger, the greater the ratio $E/J$.
Apart from this rather trivial finding, one also observes here the peculiar
dependences with the non-monotonic behavior of $T_c$.
The critical temperature may exhibit only a slight variation
with increasing $E/J$ (as it is in the case of $D/J = 1.0$), or
it may show unexpected local minima, as it is in the case
of $D/J = 2.0$ and $3.0$. Since the local minima can be located
very near to the coexistence point of the OP$_1$ and OP$_2$ phases 
(depicted as open circles),
the relevant increase of $T_c$ can be related to the OP$_2$ $\to$ OP$_1$
phase transition. Namely, the most populated $| +1/2 \rangle$ spin state 
in the OP$_2$ is replaced after this phase transition by the $| +3/2 \rangle$ 
state, which is the most occupied spin state in the OP$_1$.
The spin crossover from the low-spin $| +1/2 \rangle$ to the high-spin 
$| +3/2 \rangle$ state must lead, of course, to a slight increase of $T_c$.

At this stage, let us provide an independent check of the critical
behavior by studying thermal dependences of magnetization.
The single-site magnetization is plotted against temperature
in Fig. 5 for the biaxial anisotropy $E/J = 0.5$ and
several values of the uniaxial anisotropy $D/J$.
Fig. 5a shows a typical situation observed in the OP$_1$ phase:
the more positive the uniaxial crystal-field potential $D/J$,
the stronger the spin reduction (the lower the magnetization $m_B$)
due to the $| +3/2 \rangle \leftrightarrow | -1/2 \rangle$ spin tunneling.
In consequence of that, the total magnetization alters
from a standard Q-type dependence observed for $D \leq 0$ (see
for instance the curves for $D/J = -2.0$ and $0.0$) to a more interesting
R-type dependence, which occurs for positive uniaxial anisotropies
($D/J = 0.5$ and $0.75$). Unusual slope in the thermal
dependence of total magnetization can be related to a
more rapid thermal variation of $m_B$. In fact, on
account of the quantum fluctuations $m_B$ is thermally easier disturbed than
$m_A$ which, on the contrary, always exhibits the standard Q-type
behavior (spin-1/2 atoms are not directly affected by the biaxial
crystal-field potential $E$). Furthermore, Fig. 5b shows how the
situation changes by considering the transition toward the OP$_2$ phase.
Actually, both magnetically ordered phases OP$_1$ and OP$_2$ have the same 
internal energy (coexist together) at $D/J = \sqrt{13}/4$ when $E/J = 0.5$, 
while the OP$_2$ phase becomes more stable if $D/J > \sqrt{13}/4$.
Accordingly, Fig. 4b displays the thermal variation of sublattice magnetization 
exactly at the OP$_1$-OP$_2$ phase boundary and in the OP$_2$ phase ($D/J = 1.0$ and $1.5$).
The corresponding thermal dependences of total magnetization are plotted in the insert of
Fig. 5b. As it is apparent from these figures, the initial value of $m_B$
is suppressed from its saturation value ($m_B = 0.5$) owing to
a presence of the minority $|- 3/2 \rangle$ state.
Nevertheless, a large number of spins can be thermally excited to
the $| +3/2 \rangle$ state for $D/J$ from the vicinity of OP$_1$-OP$_2$ phase 
boundary and hence, $m_B$ rapidly increases upon heating (see the curve for $D/J = 1.0$).
As a result of this thermal excitation, the total magnetization exhibits 
N-type dependence with one compensation point in which $m_A$ and $m_B$
completely cancel out (see the insert in Fig. 5b).
Finally, even for stronger uniaxial anisotropies (e.g. $D/J = 1.5$)
the total magnetization recovers the Q-shape, since the thermal
fluctuation prefer excitations to the $| - 1/2 \rangle$ state
rather than to the $| +3/2 \rangle$ one. Such a thermal
excitations must, naturally, lower the sublattice magnetization $m_B$.

To conclude our discussion devoted to finite-temperature
properties, let us proceed to the time variation of dynamical
autocorrelation function $C_{auto}^{zz}$ as depicted in Fig. 6
for $E/J = 0.5$ and three selected values of $D/J$.
In order to enable a comparison between the displayed data at various $D/J$,
the relevant temperatures are normalized with respect to their
appropriate critical temperatures, i.e. we have defined the dimensionless temperature 
$\tau = T/T_c$ that measures a difference from the critical point
($\tau_c = 1.0$). The time variations of $C_{auto}^{zz}$ from Fig. 6a and 6b 
display the relevant changes of dynamical autocorrelation function in the OP$_1$ phase, 
while Fig. 6c shows the corresponding dependences in the OP$_2$ phase.
It can be easily understood that $C_{auto}^{zz}$ is not in general the
time-periodic function at non-zero temperatures no matter whether considering
$C_{auto}^{zz}$ in the OP$_1$, or OP$_2$ phase.
In fact, $C_{auto}^{zz}$ arises according to Eq. (\ref{eq17})
as a superposition of two harmonic oscillations with two different
angular frequencies
$\omega_{\pm} = \frac{2 J}{\hbar} \sqrt{(\frac{D}{J} \pm 1)^2 + 3 (\frac{E}{J})^2}$
and also various amplitudes.
The interference between these harmonic oscillations
gives rise to a rather complex time variation of $C_{auto}^{zz}$,
which is in general aperiodic, displaying
nodes and other typical interference effects.
The periodicity of $C_{auto}^{zz}$ at
non-zero temperatures is maintained only for some particular $E/J - D/J$
values, which retain the ratio $\omega_{+}/\omega_{-}$
to be rational, while in any other case, $C_{auto}^{zz}$ behaves
aperiodically. 

The dependences drawn in Fig. 6 nicely illustrate
also the temperature effect on the spin dynamics. It follows from
these dependences that some amplitudes are suppressed as the temperature 
increases, while another ones become more robust. Obviously,
in the high-temperature regime that amplitudes become dominant,
which coincide to the oscillation with lower angular frequency.
Contrary to this, the amplitudes arising from higher frequency oscillation
dominate in the low-temperature regime. The most miscellaneous
time variation of $C_{auto}^{zz}$ thus emerges in the vicinity of
critical temperature
($\tau \approx 1.0$), which represents an intermediate temperature range 
between the low- and high-temperature regime. However, a rather exceptional case is
displayed in Fig. 6c, where the most miscellaneous dependence
appears surprisingly at substantially lower temperature ($\tau = 0.25$)
rather than the critical one ($\tau_c = 1.0$). When looking back to the
thermal variation of magnetization depicted in Fig. 5b, one finds
a feasible explanation for this striking behavior. It turns
out that the temperature ($\tau \approx 0.25$) of the most miscellaneous 
time variation of $C_{auto}^{zz}$ coincides with the temperature $k_B T/J \approx 0.1$, 
at which the most robust spin excitation to the $| +3/2 \rangle$
can be observed. In addition to the {\it allowable} $|+ 1/2 \rangle$ and
$|-3/2 \rangle$ states, a large number of the spin-3/2 atoms is therefore thermally
excited to the $|+ 3/2 \rangle$ spin state. This observation would
suggest that the thermal excitations can basically modify the spin dynamics as well.

\section{Concluding remarks}

In this article, the exact solution of the mixed spin-1/2 and spin-3/2
Ising model on honeycomb lattice is presented and discussed in detail.
The particular attention has been focused on the effect of
uniaxial and biaxial crystal-field anisotropies acting on the
spin-3/2 atoms. As it has been shown, a presence of the biaxial
anisotropy significantly modifies the magnetic behavior of the system under 
investigation. It turns out that already a small amount of
the biaxial anisotropy raises a non-trivial spin dynamics and basically
influences the thermodynamic properties, as well.

The most striking finding to emerge here constitutes an exact
evidence of the spin tunneling between the $| \pm 3/2 \rangle$
and $| \mp 1/2 \rangle$ states in two different magnetically ordered phases
OP$_1$ and OP$_2$, respectively. Macroscopically, the tunneling
effect decreases the critical temperature of the magnetically
ordered phases and appreciably suppresses the magnetization
of spin-3/2 sublattice. This quantum reduction appears apparently 
due to the local quantum fluctuations arising from
the biaxial crystal-field potential.

There is an interesting correspondence between the model
described by the Hamiltonian (\ref{eq1}) and a similar model with
a local transverse magnetic field $\Omega$ acting on the spin-3/2 atoms only \cite{jascur}. 
However, similarity in their actual properties is not accidental, in fact, when neglecting 
the uniaxial crystal-field potential $D$ in Hamiltonian
(\ref{eq1}), an effective mapping $E \leftrightarrow \Omega$ ensures the equivalence
between both the models. Since this mapping is not related to the magnetic
structure in any fashion, the appropriate correspondence can be extended 
to the several lattice models. It is therefore valuable to mention
that magnetic properties of the models with a local transverse field
become a subject matter of many theoretical works \cite{tim}.
Apparently, the magnetic behavior of these systems should completely resemble that
one of their counterparts with the biaxial crystal-field potential.

Finally, let us turn back to the origin of biaxial anisotropy. Uprise
of this anisotropy term in the mixed-spin honeycomb lattice is, namely,
closely associated with at least a small lattice distortion.
To simplify the situation, the proposed Hamiltonian (\ref{eq1})
accounts for the biaxial crystal-field anisotropy, while a difference
between exchange interactions in the different spatial directions
has been for simplicity omitted. Nevertheless, the developed procedure
can be rather straightforwardly generalized to an anisotropic model accounting 
for the different interactions along various spatial directions. Moreover, the 
biaxial anisotropy can be even considered as an arbitrary
function (linear, quadratic, exponential, logarithmic, ...) of the ratio between
appropriate interaction parameters. Hence, it would be very interesting to find 
out whether such a system is instable toward the spontaneous lattice
distortion caused by the spin-Peierls phenomenon. In this
direction continues our next work.

\section{Appendix}

An explicit form of the coefficients $K_5$ and $K_6$ is given by:
\begin{eqnarray}
K_5 (x,y,z) = G (x,y,z)/\Phi_1, \quad \mbox{and} \quad 
K_6 (x,y,z) = G (x,y,z)/\Phi_2, \nonumber
\end{eqnarray}
where the function $G(x,y,z)$ is defined as follows:
\begin{eqnarray}
G(x,y,z) &=& \frac94[W_1(x)H_1(x,z)  + W_1(y)H_1(y,-z)] + 
             \frac14[W_2(x)H_2(x,z) \nonumber \\ &+& 
             W_2(y)H_2(y,-z)] - [W_3(x)H_3(x,z) + W_3(y)H_3(y,-z)], \nonumber \\
W_1 (x) &=& \biggl \{ x^2 + E^2 \Bigl[ 
    1 + 2 \cos(2t \sqrt{x^2 + 3E^2}/\hbar) \Bigr] \biggr \}/
                 (x^2 + 3 E^2),     \nonumber \\   
W_2 (x) &=& \biggl \{ x^2 - 3E^2 \Bigl[ 
            1 - 2 \cos(2t \sqrt{x^2 + 3E^2}/\hbar)\Bigr] \biggr \}/ 
                 (x^2 + 3 E^2),     \nonumber \\  
W_3 (x) &=& \biggl \{ \sqrt{3} E x \Bigl[ 
            1 - \cos(2t \sqrt{x^2 + 3E^2}/\hbar)\Bigr] \biggr \}/  
                 (x^2 + 3 E^2),     \nonumber \\ 
H_1 (x,y) &=& \exp(- 5 \beta D/4 - \beta y/2) \Bigl[
              \cosh(\beta \sqrt{x^2 + 3 E^2}) \nonumber \\
          &-&  x \sinh(\beta \sqrt{x^2 + 3 E^2})/\sqrt{x^2 + 3 E^2} \Bigr],  \nonumber \\ 
H_2 (x,y) &=& \exp(- 5 \beta D/4 - \beta y/2) \Bigl[
              \cosh(\beta \sqrt{x^2 + 3 E^2}) \nonumber \\
          &+& x \sinh(\beta \sqrt{x^2 + 3 E^2})/\sqrt{x^2 + 3 E^2} \Bigr],  \nonumber \\ 
H_3 (x,y) &=& \exp(- 5 \beta D/4 - \beta y/2) 
              \sqrt{3}E \sinh(\beta \sqrt{x^2 + 3 E^2})/\sqrt{x^2 + 3 E^2}.  \nonumber  
\end{eqnarray}

\textbf{Figure captions}

\begin{small}

\begin{itemize}
\item [Fig. 1]
The segment of a mixed-spin honeycomb lattice. The lattice positions
of the spin-1/2 (spin-3/2) atoms are schematically designated by
the full (open) circles, the solid lines label interactions
between nearest neighbors. The dashed lines represent the
effective interactions between three outer spin-1/2 atoms
arising after performing the mapping (\ref{eq9}) at $k$th site.
\item [Fig. 2]
a) The ground-state phase diagram in the $E/J-D/J$ plane;
b) The single-site sublattice magnetization $|m_A|$ (dotted line)
   and $m_B$ (dashed line), the total single-site magnetization
   $|m|$ (solid line) and the quadrupolar momentum $\eta$ (solid line)
   as a function of the biaxial anisotropy $E/J$ at $T = 0$ and $D/J = 1.0$.
\item [Fig. 3]
The time variation of dynamical autocorrelation function
$C_{auto}^{zz}$ at very low temperature ($k_B T/J = 0.001$)
close to the ground state:
a) for $D/J = 0.0$ and various $E/J$;
b) for $E/J = 0.5$ and various $D/J$;
Time axis is scaled in $\hbar/J$ units.
\item [Fig. 4]
a) The dependence of critical temperature on the uniaxial anisotropy
$D/J$ for several values of biaxial anisotropies $E/J$;
b) The dependence of critical temperature on the biaxial anisotropy
$E/J$ for several values of uniaxial anisotropies $D/J$.
Solid (broken) lines correspond to critical temperatures
of the OP$_1$ (OP$_2$) phase. Open circles denote the critical
temperatures for a such particular case,
when both the ordered phases OP$_1$ and OP$_2$
coexist in the ground state (see the text).
\item [Fig. 5]
The thermal dependences of single-site magnetization for
$E/J = 0.5$ and:
a) $D/J = -2.0$, $0.0$, $0.5$ and $0.75$;
b) $D/J = \sqrt{13}/4$, $1.0$ and $1.5$.
The dotted (dashed) lines stand for the sublattice magnetization
$|m_A|$ ($m_B$), the solid lines for the total single-site
magnetization $|m|$. Fig. 5b shows the temperature dependences 
of sublattice magnetization $|m_A|$ and $m_B$ only, the insert
shows the appropriate changes of total magnetization $|m|$.

\item [Fig. 6]
The time variation of dynamical autocorrelation function $C_{auto}^{zz}$
when $E/J = 0.5$ is fixed and $D/J$ changes: $D/J = -2.0$ (upper),
$0.0$ (central) and $1.0$ (lower panel). The relevant time variations of $C_{auto}^{zz}$
are displayed at three different temperatures, which are normalized with
respect to their critical temperatures in order to get the ratio
$\tau = 0.75$, $1.0$ and $1.25$ in Fig. 6ab and respectively,
$\tau = 0.1$, $0.25$ and $1.0$ in Fig. 6c.
Time axis is scaled in the $\hbar/J$ unit.
\end{itemize}
\end{small}

\newpage

\begin{figure}
\includegraphics[width=16.5cm]{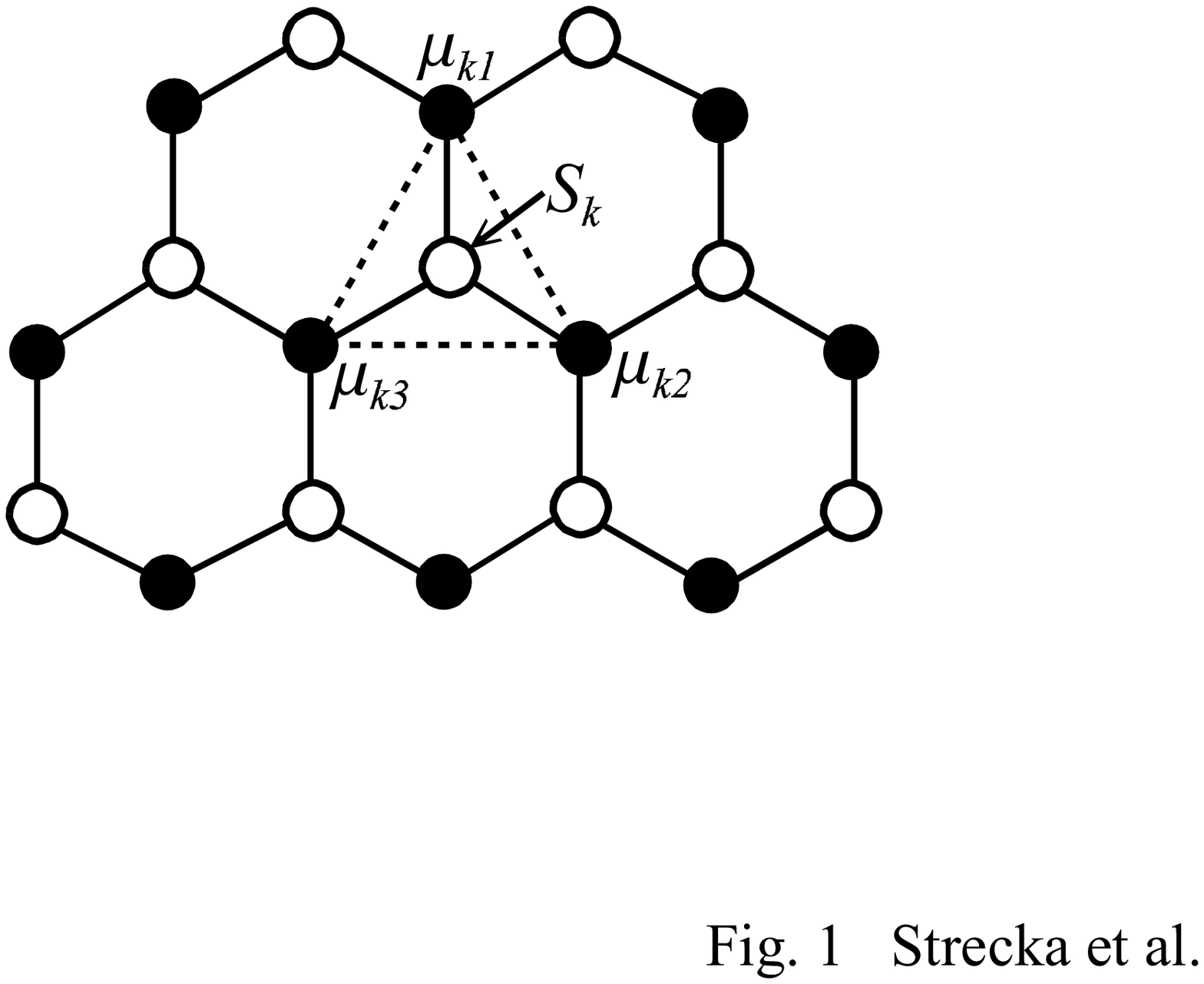}
\end{figure}

\begin{figure}
\includegraphics[width=16cm]{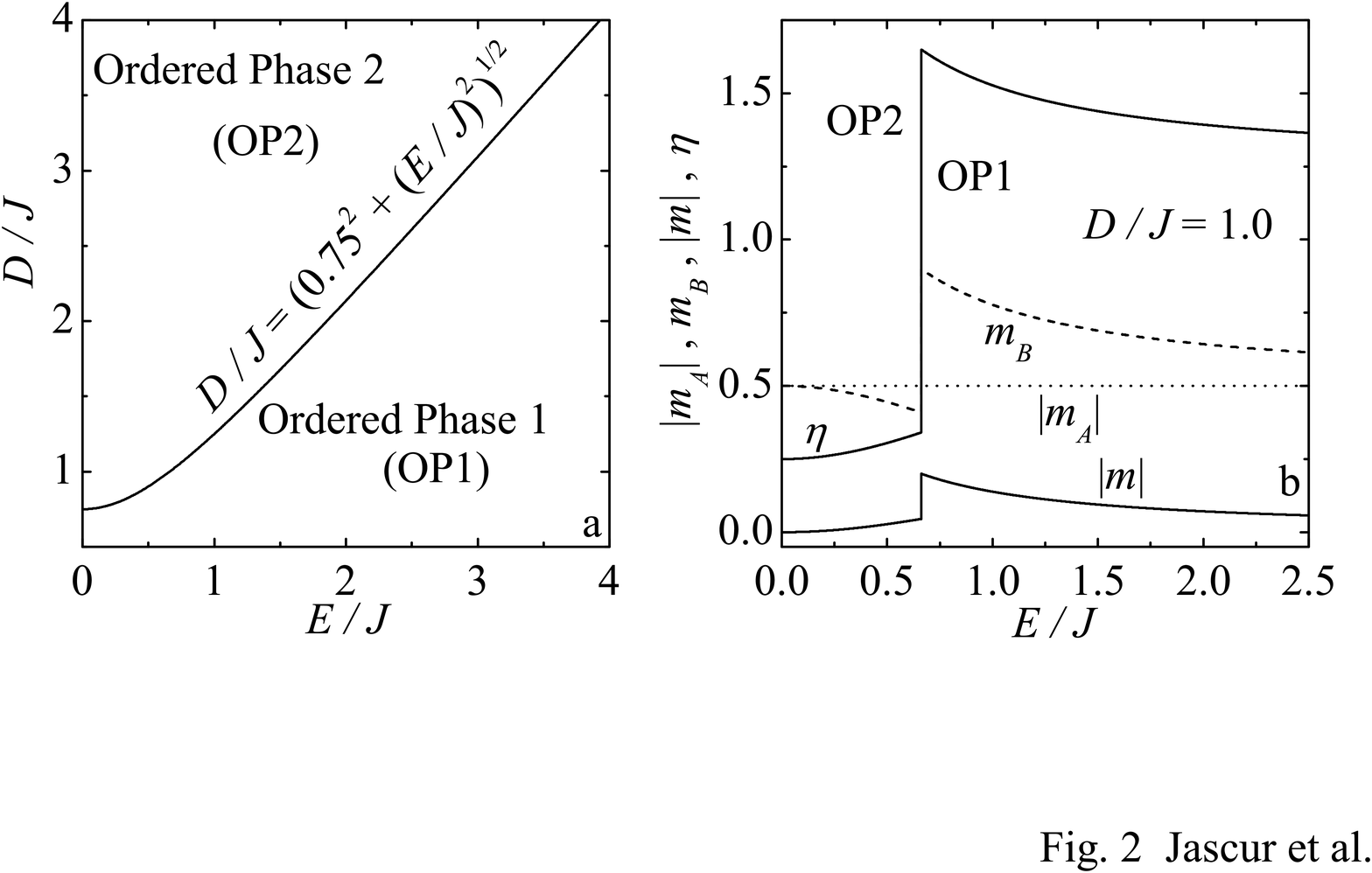}
\end{figure}

\begin{figure}
\includegraphics[width=16cm]{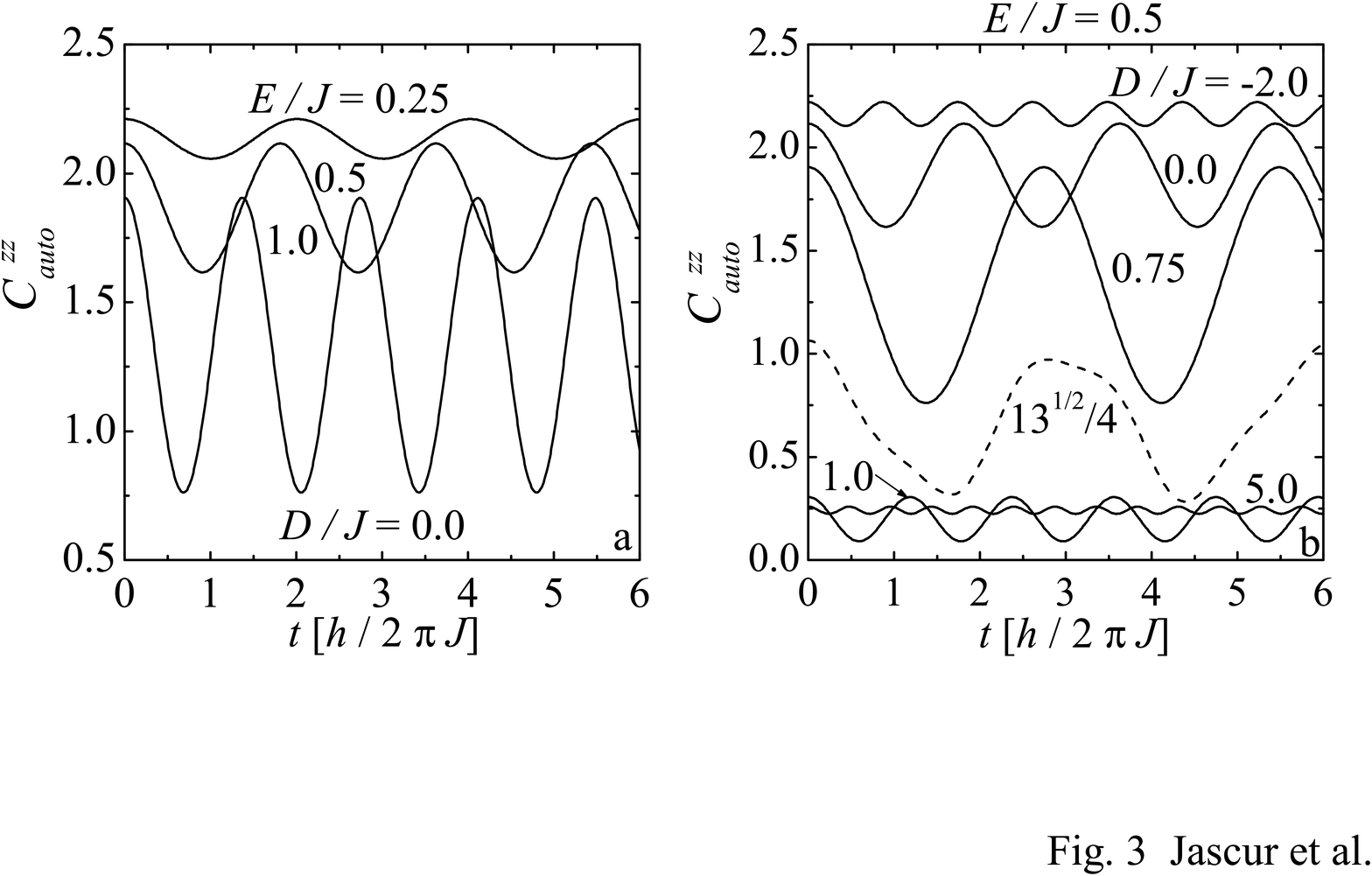}
\end{figure}

\begin{figure}
\includegraphics[width=16cm]{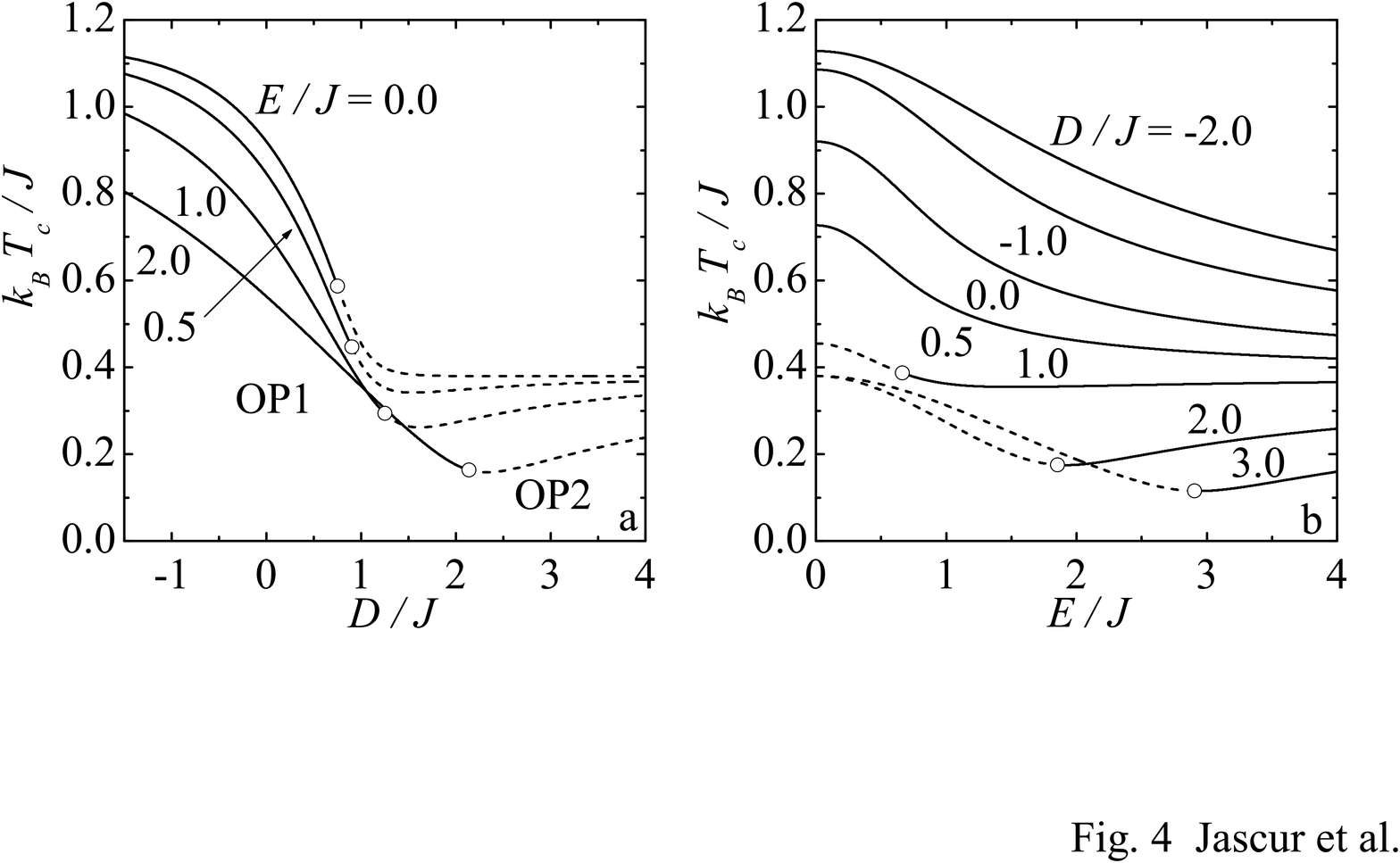}
\end{figure}

\begin{figure}
\vspace{-1cm}
\includegraphics[width=16cm]{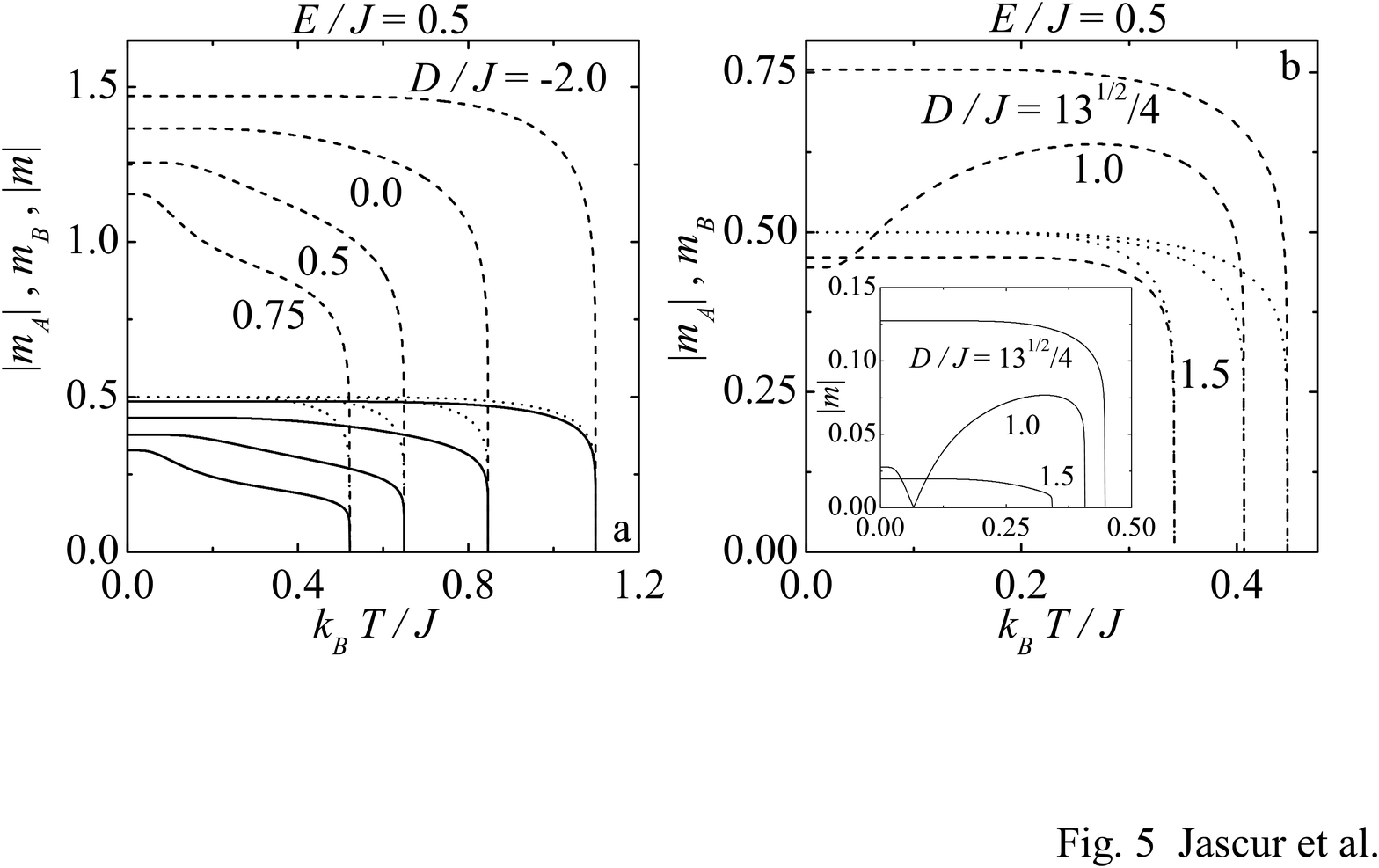}
\end{figure}

\begin{figure}
\includegraphics[width=16cm]{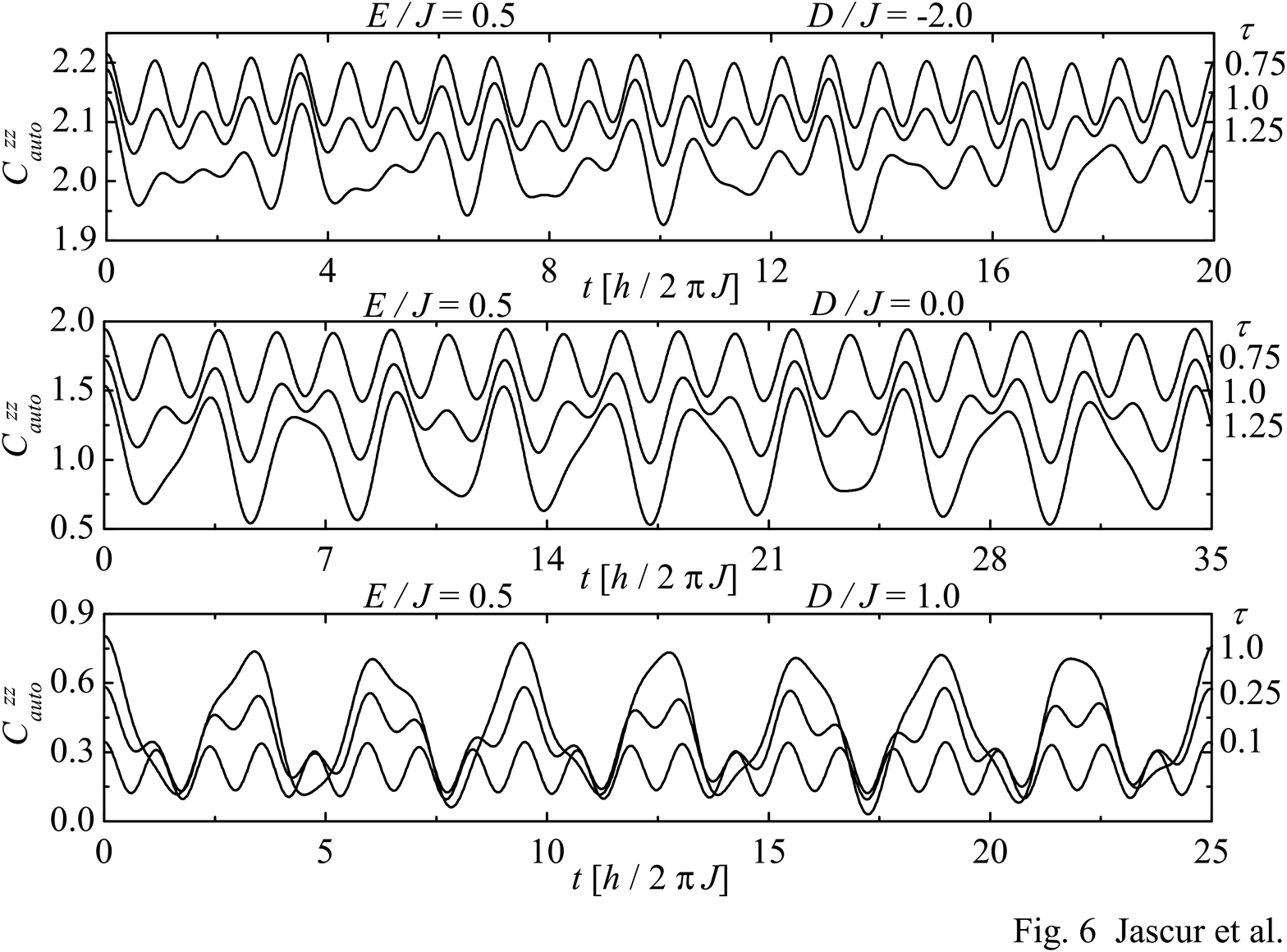}
\end{figure}

\end{document}